\documentclass{aastex62}
\usepackage{amssymb}            
\usepackage{graphicx,times}

\shorttitle{Galactic C$^{18}$O/C$^{17}$O: I. J=1-0}
\shortauthors{Zhang et al.}

\begin{document}

\title{A systematic observational study on Galactic interstellar ratio $^{18}$O/$^{17}$O:\\ I. C$^{18}$O and C$^{17}$O J=1-0 data analysis}

\correspondingauthor{JiangShui Zhang}
\email{ jszhang@gzhu.edu.cn }

\author{J.S. Zhang}
\affil{Center for Astrophysics, Guangzhou University, Guangzhou, 510006, China}

\author{W. Liu}
\affil{Center for Astrophysics, Guangzhou University, Guangzhou, 510006, China}

\author{Y.T. Yan}
\affil{Center for Astrophysics, Guangzhou University, Guangzhou, 510006, China}
\affil{Max-Planck-Institut f{\"u}r Radioastronomie, Auf dem H{\"u}gel 69, D-53121 Bonn, Germany}

\author{H.Z. Yu}
\affil{Center for Astrophysics, Guangzhou University, Guangzhou, 510006, China}

\author{J.T. Liu}
\affil{Center for Astrophysics, Guangzhou University, Guangzhou, 510006, China}

\author{Y.H. Zheng}
\affil{Center for Astrophysics, Guangzhou University, Guangzhou, 510006, China}

\author{D. Romano}
\affil{INAF, Osservatorio di Astrofisica e Scienza dello Spazio, Via Gobetti 93/3, I-40129 Bologna, Italy}

\author{Z.-Y. Zhang}
\affil{School of Astronomy and Space Science, Nanjing University, Nanjing 210093, China}

\author{J.Z. Wang}
\affil{Shanghai Astronomical Observatory, Chinese Academy of Sciences, 80 Nandan Road, Shanghai, 200030, China}

\author{J.L. Chen}
\affil{Center for Astrophysics, Guangzhou University, Guangzhou, 510006, China}

\author{Y.X. Wang}
\affil{Center for Astrophysics, Guangzhou University, Guangzhou, 510006, China}

\author{W.J. Zhang}
\affil{Center for Astrophysics, Guangzhou University, Guangzhou, 510006, China}

\author{H.H. Lu}
\affil{Center for Astrophysics, Guangzhou University, Guangzhou, 510006, China}

\author{L.S. Chen}
\affil{Center for Astrophysics, Guangzhou University, Guangzhou, 510006, China}

\author{Y.P. Zou}
\affil{Center for Astrophysics, Guangzhou University, Guangzhou, 510006, China}

\author{H.Q. Yang}
\affil{Center for Astrophysics, Guangzhou University, Guangzhou, 510006, China}

\author{T. Wen}
\affil{Center for Astrophysics, Guangzhou University, Guangzhou, 510006, China}

\author{F.S. Lu}
\affil{Center for Astrophysics, Guangzhou University, Guangzhou, 510006, China}

\begin{abstract}
The interstellar oxygen isotopic ratio of $^{18}$O/$^{17}$O can reflect the relative amount of the secular enrichment by ejecta from high-mass versus intermediate-mass stars. Previous observations found a Galactic gradient of $^{18}$O/$^{17}$O, i.e., low ratios in the Galactic center and large values in the Galactic disk, which supports the inside-out formation scenario of our Galaxy. However, the observed objects are not many and, in particular, not so many at large galactocentric distances. For this reason, we started a systematic study on Galactic interstellar $^{18}$O/$^{17}$O, through observations of C$^{18}$O and C$^{17}$O multi-transition lines toward a large sample of 286 sources (at least one order of magnitude larger than previous ones), from the Galactic center region to the far outer Galaxy ($\sim$22\,kpc). In this article, we present our observations of J=1-0 lines of C$^{18}$O and C$^{17}$O, with the 12\,m antenna of the Arizona Radio Observatory (ARO\,12\,m) and the IRAM\,30\,m telescope. Among our IRAM30\,m sample of 50 targets, we detected successfully both C$^{18}$O and C$^{17}$O 1-0 lines for 34 sources. Similarly, our sample of 260 targets for ARO\,12\,m observations resulted in the detection of both lines for 166 sources. The C$^{18}$O optical depth effect on our ratio results, evaluated by fitting results of C$^{17}$O spectra with hyperfine components (assuming $\tau_{C18O}=4\tau_{C17O}$) and our RADEX non-LTE model calculation for the strongest source, was found to be insignificant. Beam dilution does not seem to be a problem either, which was supported by the fact of no systematic variation between the isotopic ratio and the heliocentric distance, and consistent measured ratios from two telescopes for most of those detected sources. With this study we obtained $^{18}$O/$^{17}$O isotopic ratios for a large sample of molecular clouds with different galactocentric distances. Our results, though there are still very few detections made for sources in the outer Galaxy, confirm the apparent $^{18}$O/$^{17}$O gradient of $^{18}$O/$^{17}$O = (0.10$\pm$0.03)R$_{GC}$+(2.95$\pm$0.30), with a Pearson's rank correlation coefficient R = 0.69. This is supported by the newest Galactic chemical evolution model including the impact of massive stellar rotators and novae. Our future J=2-1 and J=3-2 observations of C$^{18}$O and C$^{17}$O toward the same sample would be important to determine their physical parameters (opacities, abundances, etc.) and further determine accurately the Galactic radial gradient of the isotopic ratio $^{18}$O/$^{17}$O.
\end{abstract}

\keywords{ISM: abundance -- ISM: molecules--Galaxy: evolution -- Galaxy: abundance -- radio lines: ISM}




\section{Introduction}

The evolution of metallicity in the Galactic disk depends on stellar nucleosynthesis, that is the conversion of hydrogen into heavier elements, and the ensuing contribution by ejecta to the interstellar medium (ISM, e.g., Wilson \& Rood 1994). Radial metallicity gradients in the Galactic disk are confirmed for  different objects, including stars (e.g., Xiang et al. 2017), H\,II regions (e.g., Esteban \& Garcia-Rojas 2018) and planetary nebulae (e.g., Henry et al. 2010), supporting the inside-out formation scenario for our Galaxy (Larson 1976).
Isotopic abundance ratios in the ISM are believed to be good tracers for the nucleosynthesis, the ejecta by different stars and the chemical evolution of the Milky Way (e.g., Wilson \& Rood 1994), which can be effectively measured from observations of molecular clouds and the comparison in radio bands of corresponding molecular lines of isotopologues. In particular, the $^{18}$O/$^{17}$O ratio is one of the most useful tracers of nuclear processing and metal enrichment,  readily determined from the C$^{18}$O/C$^{17}$O line intensity ratio (Wouterloot et al. 2008, Zhang et al. 2007, 2015, Li et al. 2016). In fact, it is widely accepted that $^{18}$O is synthesized by helium burning in massive stars ($M \ge 8 $~M$_\odot$) as a \emph{secondary} product of nucleosynthesis (due to the requirement of $^{14}$N as a seed nucleus), while $^{17}$O is produced through the CNO cycle (again, as a \emph{secondary} element) and ejected predominantly by longer-lived asymptotic giant branch (AGB) stars (e.g., Henkel \& Mauersberger 1993). If this is the case, the $^{18}$O/$^{17}$O ratio should abruptly increase in starbursts, following the prompt $^{18}$O release from massive stars, and decrease later on because of the delayed contribution to $^{17}$O enrichment from intermediate- and low-mass stars. In the framework of an inside-out formation for the disk of the Galaxy (Matteucci \& Francois 1989), the different nucleosynthetic paths followed by the minor oxygen isotopes should result in a mildly positive gradient of $^{18}$O/$^{17}$O along the disk (Romano et al. 2017). Actually, the nucleosynthesis of $^{18}$O and $^{17}$O in stars is more complex than the picture sketched above. Recent calculations show that at low metallicities significant amounts of \emph{primary} $^{18}$O are produced by fast-rotating massive stars (e.g., Limongi \& Chieffi 2018), while nova systems likely inject non-negligible amounts of $^{17}$O in their surroundings (e.g., Jos\'e \& Hernanz 1998, 2007). Given the complex dependence of the yields on the mass, metallicity and rotation of the stars, it becomes thus harder to predict the behavior of the $^{18}$O/$^{17}$O ratio without the aid of a self-consistent chemical evolution model. In Sect. 3.4 we examine in detail the predictions of a well-tested chemical evolution model for the Galaxy (Romano et al. 2019), when the prescriptions about $^{18}$O and $^{17}$O synthesis in stars are changed and compare these with our observational results.

Almost 40 years ago, Penzias (1981) reported one uniform $^{18}$O/$^{17}$O value of $\sim$3.5, from molecular clouds in the GC out to the Perseus arm with the galactocentric distance (R$_{\rm GC}$) of $\sim$10\,kpc. More recently, Wouterloot et al. (2008) suggested the existence of a $^{18}$O/$^{17}$O gradient along the galactic disk (R$_{\odot}$=8.5\,kpc), with average isotopic ratio of 2.88$\pm$0.11, 4.16$\pm$0.09 and 5.03$\pm$0.46 for SgrB2 in the Galactic center, sources at 4--11\,kpc and sources at $\sim$16.5\,kpc, respectively. However, this possible gradient suffers from small number statistics (18 sources), in particular, with few sources in the Galactic center and in the far outer Galaxy (with R$_{\rm GC}$$>$16\,kpc).

For this reason, we conducted a systematic observation of the interstellar isotopic ratio of $^{18}$O/$^{17}$O across the entire Galaxy (out to R$_{\rm GC}$$\sim$22\,kpc). We have performed C$^{18}$O and C$^{17}$O mapping toward molecular clouds in the Galactic center for the first time (Zhang et al. 2015) and a single-pointing pilot survey of Galactic disk molecular clouds (14 sources) with different distances with the Delingha 13.7\,m telescope (DLH\,13.7m) in Purple Mountain Observatory (Li et al. 2016). Our analysis of C$^{18}$O and C$^{17}$O (J=1-0) clearly shows a lower abundance ratio toward the GC region with respect to molecular clouds in the Galactic disk, though the ratios appear not to be uniform inside the GC region. Here we present observations of C$^{18}$O and C$^{17}$O J=1-0 toward a sample of 286 molecular clouds in the Galactic disk, including 20 sources in the far outer Galaxy (R$_{\rm GC}$$>$16\,kpc). Our sample includes star formation regions associated with IRAS sources with relatively strong CO emission (the corrected antenna temperature T$_A^{*}$$>$10\,K, selected from Wouterloot \& Brand 1989), and high-mass star forming regions (HMSFRs) whose distances have been accurately measured by maser parallax method (Reid et al. 2014). Our observations with the ARO\,12\,m and the IRAM\,30\,m are described in Sect.\,2,  the analysis of J=1-0 spectral data and corresponding results are presented and discussed in Sect.\,3, while Sect.\,4 summarizes the main results.

\section{Observations}

\subsection{IRAM\,30m observations}

The C$^{18}$O and C$^{17}$O J=1-0 lines were observed from June 29 to July 1 2016, with the IRAM 30\,m single dish telescope\footnote{The IRAM\,30m is supported by INSU/CNRS (France), MPG (Germany) and IGN (Spain).} at Pico Veleta Observatory (Granada, Spain). The center frequencies were set at 109.782182 and 112.359277\,GHz for the C$^{18}$O and C$^{17}$O, respectively, with a corresponding beam size of $\sim$23$\arcsec$ (Pagani et al. 2005). The observations were carried out in position switching mode with the off position at (-1800$\arcsec$, 0$\arcsec$) or (1800$\arcsec$, 0$\arcsec$) offset in R.A. and Dec. from the source. The Eight Mixer Receiver (EMIR) with dual-polarization and the Fourier Transform Spectrometers (FTS) backend were used, providing frequency coverage of 108--116\,GHz in the upper sideband, with a spectral resolution of 195\,kHz or $\sim$0.5\,km\,s$^{-1}$ around 110\,GHz. The typical system temperature was 170--300\,K. C$^{18}$O and C$^{17}$O lines were observed simultaneously with an rms noise at antenna temperature scale T$^{*}_{A}$ of about 10--40\,mK. The main beam brightness temperature T$_{mb}$ can be obtained from the antenna temperature multiplied by the ratio of the forward efficiency and the main beam efficiency (F$_{eff}$/B$_{eff}$$\sim$ 0.94/0.78 = 1.21).

With the IRAM\,30m we observed 50 sources of our sample. The parameters of their observations are summarized in Table\,1. The source name and its equatorial coordinates in J2000 are listed in Columns 1-3. The galactocentric distance and the heliocentric distance of each source, used telescope and targeted molecular species are listed in Columns 4-7. The Column 8 provides the rms noise of observations and, Columns 9 and 10 list the integrated line intensity and the peak temperature, respectively, obtained from Gaussian fits to the spectra. For each source on the first line are the results for C$^{18}$O, on the second line those for C$^{17}$O. In Column 11 is the abundance ratio (see details in Sect.3.1).

\subsection{ARO\,12m Observations}

 Other observations of the $J=1-0$ lines of C$^{18}$O and C$^{17}$O were also carried out using the ARO\,12\,m telescope on Kitt Peak, Tuscon, AZ, USA\footnote{The ARO\,12\,m is supported by the Department of Astronomy and Steward Observatory of the University of Arizona (USA).}, with a corresponding beam size of $\sim$64$\arcsec$ (Kelly et al. 2015). Observations were performed remotely from Guangzhou University, China, in November and December 2016, November and December 2017 and January, May and December 2018. The newer dual-polarization receiver containing ALMA Band 3 (83$\sim$116\,GHz) sideband-separating (SBS) mixers was employed. The 12 Meter millimeter autocorrelator (MAC) and the 12M Filters backends were used during our observations in 2016 \& 2017, and a new ARO Wideband Spectrometer (AROWS) backend was used in our later observations. The MAC backend used a 300\,MHz bandwidth (6144 channels), with a spectral resolution of 48.8\,kHz ($\sim$0.13\,km\,s$^{-1}$), while the 12M Filters backends used a 25.6\,MHz or 64\,MHz bandwidth (256 channels), with a spectral resolution of 100\,kHz ($\sim$0.27\,km\,s$^{-1}$) or 250\,kHz ($\sim$0.67\,km\,s$^{-1}$), respectively. The AROWS backend used a 120\,MHz bandwidth (6400 channels), with a spectral resolution of 18.75\,kHz ($\sim$0.05\,km\,s$^{-1}$). Observations were performed in position switching mode with an off-position of 30$'$. The system temperature was 80-310\,K on the antenna temperature scale T$^{*}_{A}$, with an rms noise of about 40\,mK. The main beam brightness temperature T$_{mb}$ can be determined from the antenna temperature scale by T$_{mb}$ = T$^{*}_{A}$/$\eta_{b}$, where $\eta_{b}$ is the main beam efficiency correction factor (E.g., Calahan et al. 2018), with a mean value of $\sim$0.83 during our observations.

Table\,1 presents the parameters of the observational results by ARO for the sample of 260 sources, including 24 sources observed by both IRAM\,30m and ARO\,12m telescopes. Comparisons of observation results of the same source from different telescopes are needed to evaluate calibration uncertainties and the effect of the beam size on the line ratios.


\section{Results \& Discussion}

\subsection{Spectra fitting results}

ARO\,12m detected in both J=1-0 C$^{18}$O and C$^{17}$O lines 166 sources out of 260 targets, while IRAM\,30m detected in both lines 34 sources out of 50 targets. Seven sources were detected in both lines by ARO\,12m and by IRAM\,30m. In total, 193 sources among our sample of 286 Galactic molecular clouds have been detected both in J=1-0 C$^{18}$O and in C$^{17}$O lines. Figure\,1 and Figure\,2 present the C$^{18}$O and C$^{17}$O spectra for all our detections from the ARO\,12m and the IRAM\,30m, respectively. The spectra of the targets that were not detected (or with only hints of a detection, candidates for additional observation) are shown in the Appendix.

Data were reduced using the CLASS software of the GILDAS packages (e.g. Guilloteau \& Lucas 2000). Baselines subtraction was done for all detected lines. Line parameters were obtained from Gaussian fits (see green lines in Figure 1 and Figure 2) to C$^{18}$O and C$^{17}$O spectra. Sources that displayed C$^{17}$O spectra with hyperfine structure (hfs) features (see detail in sect.\,3.3), due to the interaction of the
electric quadrupole and magnetic dipole moment of the $^{17}$O nucleus (Frerking \& Langer 1981), were analyzed using hfs fits
("method'' command in CLASS), with fixed velocity shifts of 0.555, 0 and -2.715\,km\,s$^{-1}$ and relative intensity of 2.1, 4.1 and 3.3 for the
three hfs components (Frerking \& Langer 1981).
Sources with low S/N ($\leq$3) C$^{17}$O spectra (WB89\,018, WB89\,179, WB89\,250, WB89\,258, WB89\,262 and WB89\,266), were not used for our later analysis. In addition, we note that many sources show absorption features in their spectra, due to the off-positions not being emission-free. In 33 of these, the absorption is very close in velocity to the emission line and may affect it and consequently the line ratios. Thus these sources were also not used for our later analysis.

Since the column density is proportional to $\nu^{-2}$ times the integrated line intensity for clouds filling the beam, the abundance ratio C$^{18}$O/C$^{17}$O can be obtained from the integrated intensity ratio C$^{18}$O/C$^{17}$O times the factor $(\nu_{\rm C17O}/\nu_{\rm C18O})^2$ = 1.047 due to the frequency difference of two lines (e.g., Linke et al. 1977, Zhang et al. 2015). Our spectral fitting results, including the peak value (T$_{peak}$), the integrated intensity of C$^{18}$O and of C$^{17}$O with their uncertainties, and the abundance ratios with their uncertainty,
are also presented in Table\,1 (Columns 9-11).

\clearpage
\subsection{Possible effects on abundance ratios}

\subsubsection{Observational effects:}

The linear beam size is different for sources at different distances. A larger beam size of sources at larger distance may include more relatively diffuse low density gas, which could affect the isotope ratio results. And it may also bias to brighter and more massive sources, with possible systematically higher C$^{18}$O opacities, which would underestimate the ratios of C$^{18}$O/C$^{17}$O (Wouterloot et al. 2008). In order to assess possible beam dilution effects on our ratios, we plotted the isotopic ratio against the heliocentric distance in Figure\,3. It shows no systematic dependence between the ratio and the distance, which means that any observational bias related to the beam dilution is not significant for our ratio results.

Since our data come from two different telescopes ARO\,12m and IRAM\,30m, with a beam area difference by a factor of $\sim$8, thus we investigate the beam dilution effects between data from the two antennas. We compared the spectra for the 7 sources detected in both lines with ARO\,12m and IRAM\,30m (the sources span a Galacocentric distance of $\sim$5--13\,kpc). All individual spectral lines from the same source (with the exception of WB89\,191) show a stronger signal in IRAM\,30m observation when compared to ARO\,12m one (Figure\,4). This is due to the fact that the sizes of our sources are smaller than the beam size and the dilution in the beam area is not negligible (larger for ARO\,12m as it has larger beam). For WB89\,191, the T$_{mb}$ of its IRAM\,30m spectra is a little smaller than that of its ARO\,12m spectra. Assuming that pointing uncertainties affect only marginally our observations, this may suggest that this source is extended (with a bigger size than both beams) and nonuniform with a clumpy structure (covered by ARO\,12m, but beyond IRAM\,30m beam). Therefore, it is an interesting candidate for further mapping observations, probing its large scale extension and looking for small clumps.

The abundance ratio of the sources detected by both telescopes were also compared (with the exception of WB89\,477, because its ARO C$^{17}$O spectrum shows only one very narrow feature and no hfs features, despite they are quite evident in the spectrum taken by IRAM\,30m, see (Figure\,4). These sources show consistent ratio results from IRAM\,30m and ARO\,12m within the uncertainty ranges (Table\,2), which may reflect the beam dilution effect is not significant on our ratio results. Since the two lines are observed separately by ARO\,12m, while they are observed simultaneously by IRAM\,30m, calibration uncertainties may arise from variable telescope efficiency or pointing uncertainties (Wouterloot et al. 2008). As the ratios computed from the data taken by the two telescopes are compatible, we believe that there are no significant calibration uncertainties on the estimates of the line ratio.

\subsubsection{Chemical and physical processes:}

The isotopic ratio from corresponding molecular isotopologue ratio might be affected by chemical and/or physical process, including chemical fractionation, isotope-selective photo-dissociation, opacity effect. The chemical fractionation of oxygen is negligible due to its high first
ionization potential of $\sim$12.07\,keV, thus efficiently inhibiting O$^+$ charge exchange reactions (Langer et al. 1984). Although the C$^{18}$O line is normally optical thin, it becomes optical thick when its column density is larger than 10$^{16}$\,cm$^{-2}$ (Wouterloot et al. 2005). Its large optical depth could saturate the C$^{18}$O line, which leads to underestimating the real abundance ratio C$^{18}$O/C$^{17}$O. On the other hand, large opacity of C$^{18}$O would be less affected by photo-dissociation through self-shielding against UV radiation than the rarer C$^{17}$O, i.e., isotope-selective photo-dissociation effect. This would increase the measured intensity ratio C$^{18}$O/C$^{17}$O.

To estimate the influence of opacity effect in C$^{18}$O, we determine directly the optical depth of C$^{17}$O by fitting its hfs components.
We fit those good quality C$^{17}$O spectra that display hfs features, including 11 sources by IRAM\,30m (G034.04+00.15, G035.14-00.73, G059.78+00.06, WB89\,022, WB89\,081, WB89\,123, WB89\,171, WB89\,182, WB89\,184,WB89\,194, WB89\,448) and 13 sources by ARO\,12m (WB89\,195, WB89\,252, WB89\,257, WB89\,261, WB89\,170, WB89\,210, WB89\,307, WB89\,349,  WB89\,463, WB89\,522, G078.12+03.63, G079.73+00.99, G160.14+03.15). The C$^{17}$O optical depth is not larger than 0.1 for all sources except WB89\,170 and G079.73+00.99, which have an optical depth of 0.16$\pm$0.18 and 0.22$\pm$0.14, respectively. With a assumption of the optical depth of C$^{18}$O to be four times the one of C$^{17}$O (e.g., Wouterloot et al. 2008), the optical depth of C$^{18}$O of WB89\,170 and G079.73+00.99 is about 0.64 and 0.88, which increases the ratio of C$^{18}$O/C$^{17}$O by 25\% and 35\%, respectively. For all other sources, the C$^{18}$O optical depth is small ($\leq$0.4) and the influence of opacity effect is not significant.

In addition, the RADEX non-LTE (Local Thermodynamic Equilibrium) radiative transfer model (Van der Tak et al. 2007) was used to calculate the optical depth for the strongest source (G034.30+00.15) in our sample. Since its size of $\sim$2$\arcmin$ (Sanhueza et al. 2010) is larger than the ARO\,12m beam, we can safely neglect the filling factor effect and use in the model the peak intensity, 8.10\,K, and the line-width, 5.56\,km\,s$^{-1}$ for its C$^{18}$O spectra. Additionally, we use the new set of collision rate coefficients calculated by Yang et al. (2010), the kinetic temperature range of 10\,K$\sim$50\,K and the volume density range of 10$^{4}$\,cm$^{-3}$$\sim$10$^{6}$\,cm$^{-3}$. Figure\,5 presents the calculated optical depth against the kinetic temperature, for different densities. It shows clearly that the calculated optical depth is not affected by the volume density. The kinetic temperature of $\sim$35\,K, estimated by Myers et al. (1996), is obtained from the fit of a simple analytic model to spectral lines of multiple molecules (CS, C$^{18}$O, CO etc.). At this temperature, the optical depth of C$^{18}$O is $\sim$0.3, which implies a correction of $\sim$10\% on the ratio of C$^{18}$O/C$^{17}$O for the strongest source. This supports our results of the hfs-fit method, i.e., for most of our sources, the optical depth should not have a significant effect on the determination of the line ratios.


\subsection{Galactic interstellar $^{18}$O/$^{17}$O gradient}

As mentioned in Sect.\,3.1, the 33 sources with absorption features in their spectra are excluded from our analysis. Thus we got the isotopic ratio $^{18}$O/$^{17}$O for 154 sources among our sample. Ratios are plotted against their galactocentric distances in Figure\,6a (R$_{\odot}$=8.5\,kpc). Filled black triangles and red circles are the results from our ARO\,12m and IRAM\,30m measurements, respectively. Previous measurements by the DLH\,13.7m of the Purple Mountain Observatory are presented as green empty squares. Similar to Wouterloot et al. (2008),  we report a low $^{18}$O/$^{17}$O of $\sim$3 for the Galactic center region and a large average value of $\sim$4-5 in the outer Galaxy, with an almost constant value of $\sim$3.5 from $\sim$3\,kpc to 12\,kpc. The trend is obvious for either the whole sample and for each sub-sample, i.e., the ratio of $^{18}$O/$^{17}$O increases with the galactocentric distance R$_{\rm GC}$, though the scatter is large and only a few detections are made for sources in the outer Galaxy (5 sources ranging 16\,kpc to 20\,kpc, while 4 sources around 16.5\,kpc in Wouterloot et al. 2008).

In order to provide a robust estimate of the gradient in terms of the galactocentric distance, we analyze the subsample with HMSFR data, which has fairly accurate distance values, thanks to the measurements of their trigonometric parallaxes and proper motions for masers (Reid et al. 2014). Results are plotted in Figure\,6b and the increasing trend of $^{18}$O/$^{17}$O is evident. In order to show the trend more clearly, we average our data in bins of 1\,kpc in galactocentric distance and compute weighted average ratios in each bin. Both unweighted and weighted least squares fits (the solid and dashed line, respectively) are made and presented in Figure\,7. With regard to the unweighted fit, the weighted fit line tends to be flat, dominated by the low ratios with small uncertainties in the inner Galaxy. Following Wouterloot et al. (2008) and Yan et al. (2019) we prefer the unweighted fit, which results in a radial gradient of $^{18}$O/$^{17}$O is: $^{18}$O/$^{17}$O = (0.10$\pm$0.03)R$_{GC}$+(2.95$\pm$0.30), with a Pearson's rank correlation coefficient R = 0.69.

Either for our whole sample or the HMSFR sample alone, the radial gradient of $^{18}$O/$^{17}$O is apparent, though the values are much scattered and only a few measurements are available for the far outer Galaxy. Theoretical models for galactic chemical evolution (GCE) are helpful for us to understand the evolution of isotopic ratios in the Galaxy. Recently, new GCE models have been published that track the cosmic evolution of the CNO isotopes in the interstellar medium (ISM) of galaxies, yielding powerful constraints on their stellar initial mass function (IMF, Romano et al. 2017, 2019; Zhang et al. 2018). The curves in Figure 6 show the theoretical $^{18}$O/$^{17}$O gradient across the Milky Way disk at the present time predicted by models MWG-10 and MWG-12 of Romano et al. (2019). Furthermore, we run model MWG-12 with no novae (hereafter MWG-12nn), which is the same as model MWG-12, but with the nova contribution to the $^{17}$O synthesis set to zero. The models adopt the yields of Karakas (2010) for single low- and intermediate-mass stars (1--6 M$_\odot$), the yields of Doherty et al. (2014a, b) for super-AGB stars (with masses in the range 7--9\,M$_\odot$) and the yields of Limongi \& Chieffi (2018) for massive stars ending as core-collapse supernovae ($M > 10$ M$_\odot$). The possibility that massive stars are (MWG-12 and MWG-12nn, see solid and dash-dot curves in Figure 6) or are not (MWG-10, dot curves in Figure 6) rotating fast below a critical metallicity threshold ([Fe/H]$< -$1; see Romano et al. 2019 for details) is explored. Massive fast rotators can produce large amounts of $^{18}$O at low metallicity, thanks to the activation of a primary $^{18}$O production channel by rotation (see discussion in Limongi \& Chieffi 2018, and references therein). Therefore, when fast rotators are implemented in the chemical evolution code, higher $^{18}$O/$^{17}$O ratios are predicted in the outer, metal-poor Galactic disk, than in the case without fast rotating stars. The resulting $^{18}$O/$^{17}$O gradient is overall positive, while it has a more complex behavior if stellar rotation is not considered (it is flattening in the innermost and outermost Galactic disk regions, and presents a negative trend going from R$_{GC}$=8\,kpc to R$_{GC}$=12\,kpc). This is due to the non-linear dependence of the yields on the mass and metallicity of the stars. If the contribution from novae (see Jos\'e \& Hernanz 2007; Romano et al. 2017, and references therein) is taken into account, the positive slope of the gradient is preserved, but the ratios are slightly lower (because of the negligible synthesis of $^{18}$O compared to the significant $^{17}$O production from these stellar factories). As stressed by Romano et al. (2017), however, the inclusion of nova nucleosynthesis in chemical evolution models is affected by many uncertainties, so the results must be taken with care. In particular, we assume that all novae eject the same amount of $^{17}$O during their lifetime, independently of the metallicity of the progenitor system. It is unclear at present if a metal-dependent nova yield should be considered. For the purpose of the present study, it suffices to show that novae may have a significantly impact the $^{18}$O/$^{17}$O gradient on a Galactic scale and, therefore, they should not be neglected in GCE studies (see model MWG-12 and model MWG-12nn predictions in Fig.6). Overall, if the unweighted fit to the bins of data (the solid line in Figure\,7) is considered,the agreement between the data and the model predictions including both novae and rotating stars (below [Fe/H]=$-$1) is satisfactory (the solid curve in Figure\,7), with a deviation of $\leq$15\%. The more precise the $^{18}$O/$^{17}$O gradient determination, the tighter the constraints we can impose on novae contribution to the synthesis of CNO isotopes.

Our coming J=2-1 and J=3-2 observations of C$^{18}$O and C$^{17}$O toward our sample and theoretical modeling works would be important to determine their physical parameters (opacities, abundances, etc.) and further determine accurately the Galactic radial gradient of the isotopic ratio $^{18}$O/$^{17}$O.


\section{Summary}

To investigate how the oxygen isotopic ratio of $^{18}$O/$^{17}$O varies across our Galaxy, we are performing one systematic observations of multi-transition lines of C$^{18}$O and C$^{17}$O toward a large sample of 286 molecular clouds in the Galactic disk, covering from the the Galactic center region to the far outer Galaxy ($\sim$22\,kpc). In this work, we presented and analyzed our observational data of J=1-0 lines of C$^{18}$O and C$^{17}$O, obtained with the ARO\,12m and the IRAM\,30\,m telescopes. The IRAM\,30m detected successfully both C$^{18}$O and C$^{17}$O 1-0 lines toward 34 sources out of 50 targets. And 166 of 260 targets were detected with both lines by the ARO\,12m telescope. In order to assess possible effects on our results from different beam sizes, 24 sources were observed by both telescopes and 7 out of them were detected with both C$^{18}$O and C$^{17}$O 1-0 lines. Our results include:

1) The reliable abundance ratios of C$^{18}$O/C$^{17}$O reflecting the $^{18}$O/$^{17}$O isotopic ratio are obtained firstly for a large sample of 154 sources, according to measured J=1-0 line intensity ratios of C$^{18}$O and C$^{17}$O. An observational bias due to the beam dilution is not apparent, based on the fact that no systematic variation appears between the isotopic ratio and the heliocentric distance. The beam dilution effects between data from two telescopes are found also to be insignificant, since consistent measured ratios from two telescopes are given for most of those sources detected by both ARO12\,m and IRAM\,30m.

2) For those sources with C$^{17}$O hfs features, our fitting obtained the C$^{17}$O optical depth and further got the C$^{18}$O optical depth , assuming $\tau_{C18O}=4\tau_{C17O}$. All $\tau_{C18O}$ values are found to be less than 1. Moreover, our RADEX non-LTE model calculation results show that the optical depth of the strongest source among our sample is also small, with a value of $\sim$0.3 at a kinetic temperature of 35\,K. These results support that the opacity effect should not have a great influence on our estimates of the isotopic ratio $^{18}$O/$^{17}$O.

3) Based on a large sample covering sources from the Galactic center to the far outer Galaxy, our results confirm the Galactic radial gradient of $^{18}$O/$^{17}$O proposed by previous works, though the scatter is large and more $^{18}$O/$^{17}$O measurements of sources in the outer Galaxy are still needed. This is also apparent for the subsample of high-mass star forming region, whose targets have an accurate galactocentric distance derived from the trigonometric parallax. A unweighted least-squares linear fit to data averaged in 1 kpc-wide bins in R$_{GC}$ gives $^{18}$O/$^{17}$O = (0.10$\pm$0.03)R$_{GC}$+(2.95$\pm$0.30), with a Pearson's rank correlation coefficient R = 0.69. This is basically consistent with predictions of the GCE model, including rotating star yields and $^{17}$O ejecta from novae, as well as the nucleosynthesis from single rotating high, intermediate and low-mass stars.

\begin{acknowledgements}
 This work is supported by the Natural Science Foundation of China (No. 11590782, 11590783, 11473007). DR and ZYZ acknowledge support from the International Space Science Institute (ISSI, Bern, CH) and the International Space Science Institute-Beijing (ISSI-BJ, Beijing, CN) through the funding of the team ``Chemical abundances in the ISM: the litmus test of stellar IMF variations in galaxies across cosmic time". We thank the operators and staff at both ARO\,12m and IRAM\,30m stations for their assistance during our observations. Many thanks to the referee for his careful reading and good comments, and to Dr. D. Bastieri for his help on English improvement of the manuscript.

\end{acknowledgements}


\setcounter{figure}{0}
\begin{figure}
\plotone{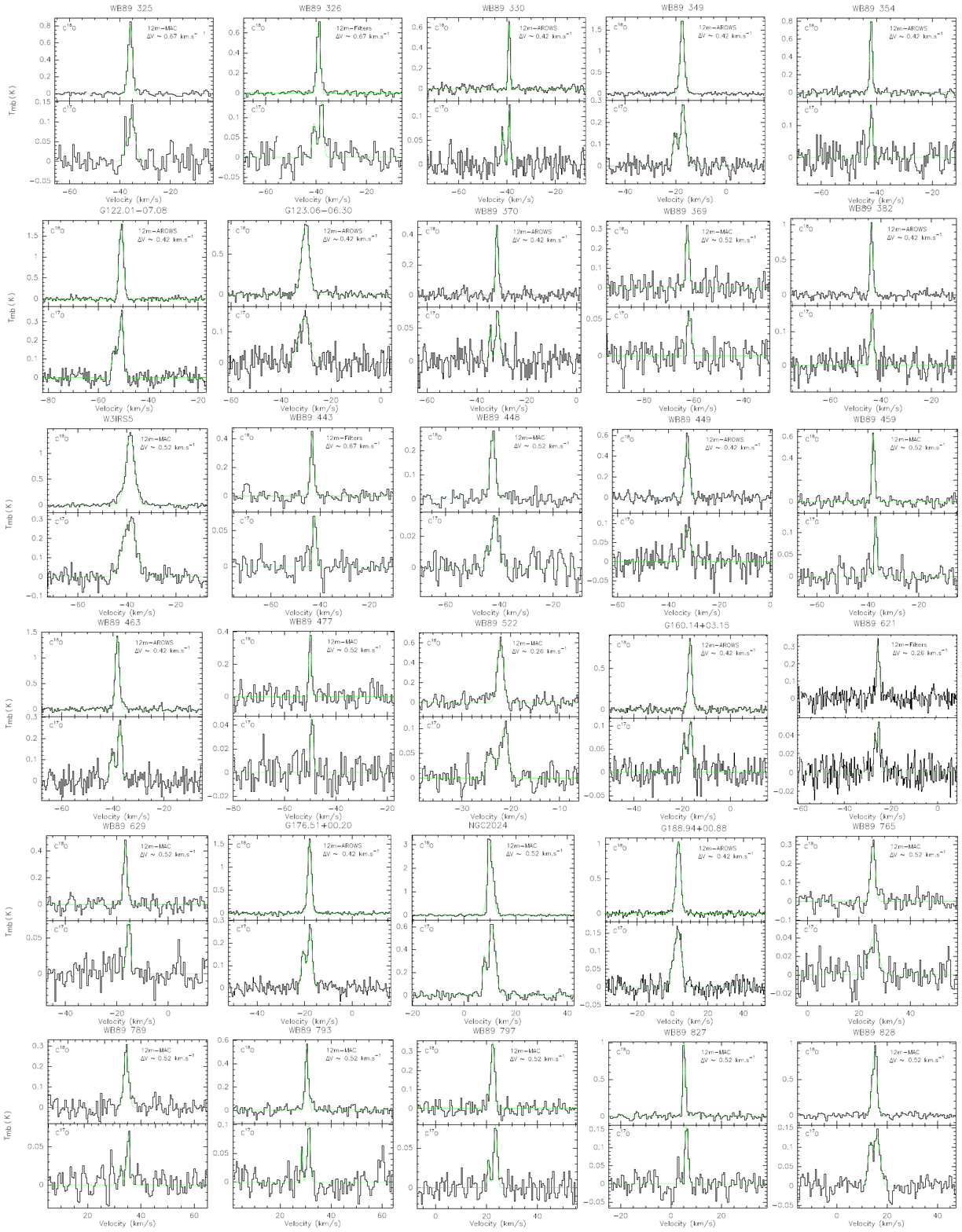}
\caption{The ARO\,12m spectra of C$^{18}$O (upper panels) and C$^{17}$O (lower panels) with green fit lines of the 166 sources in both isotopomers.}
\end{figure}

\setcounter{figure}{0}
\begin{figure}
\plotone{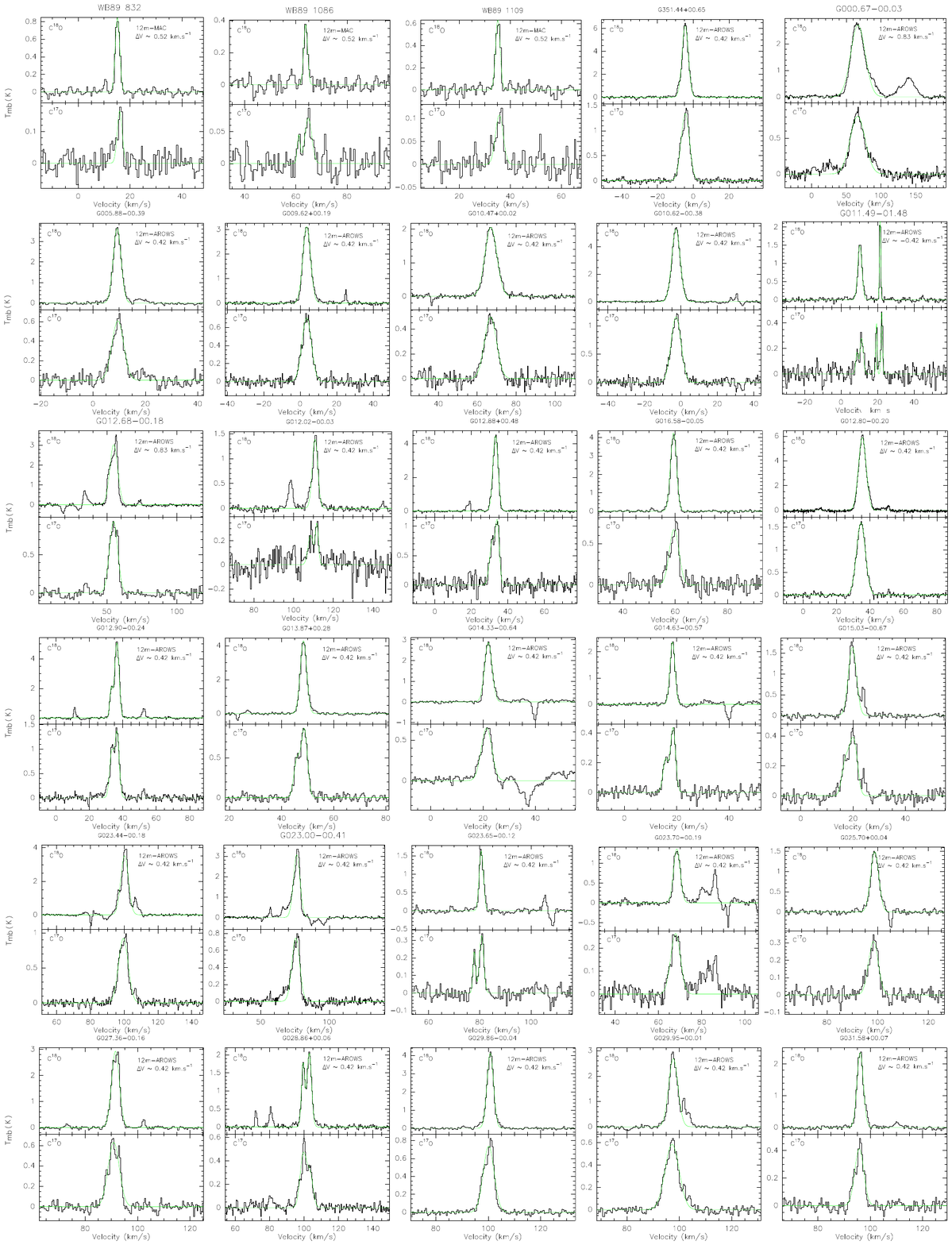}
\caption{continued}
\end{figure}

\setcounter{figure}{0}
\begin{figure}
\plotone{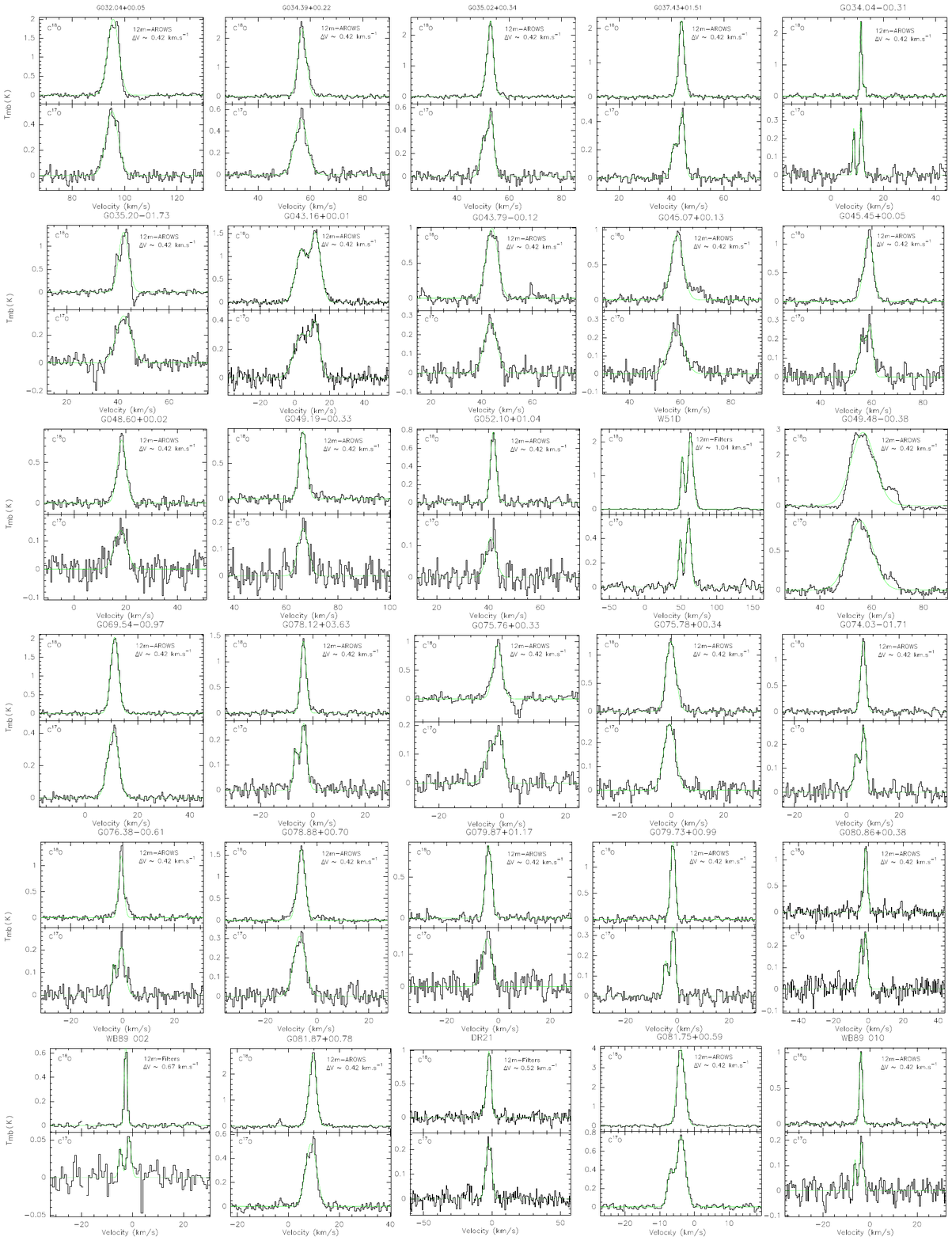}
\caption{continued}
\end{figure}

\setcounter{figure}{0}
\begin{figure}
\plotone{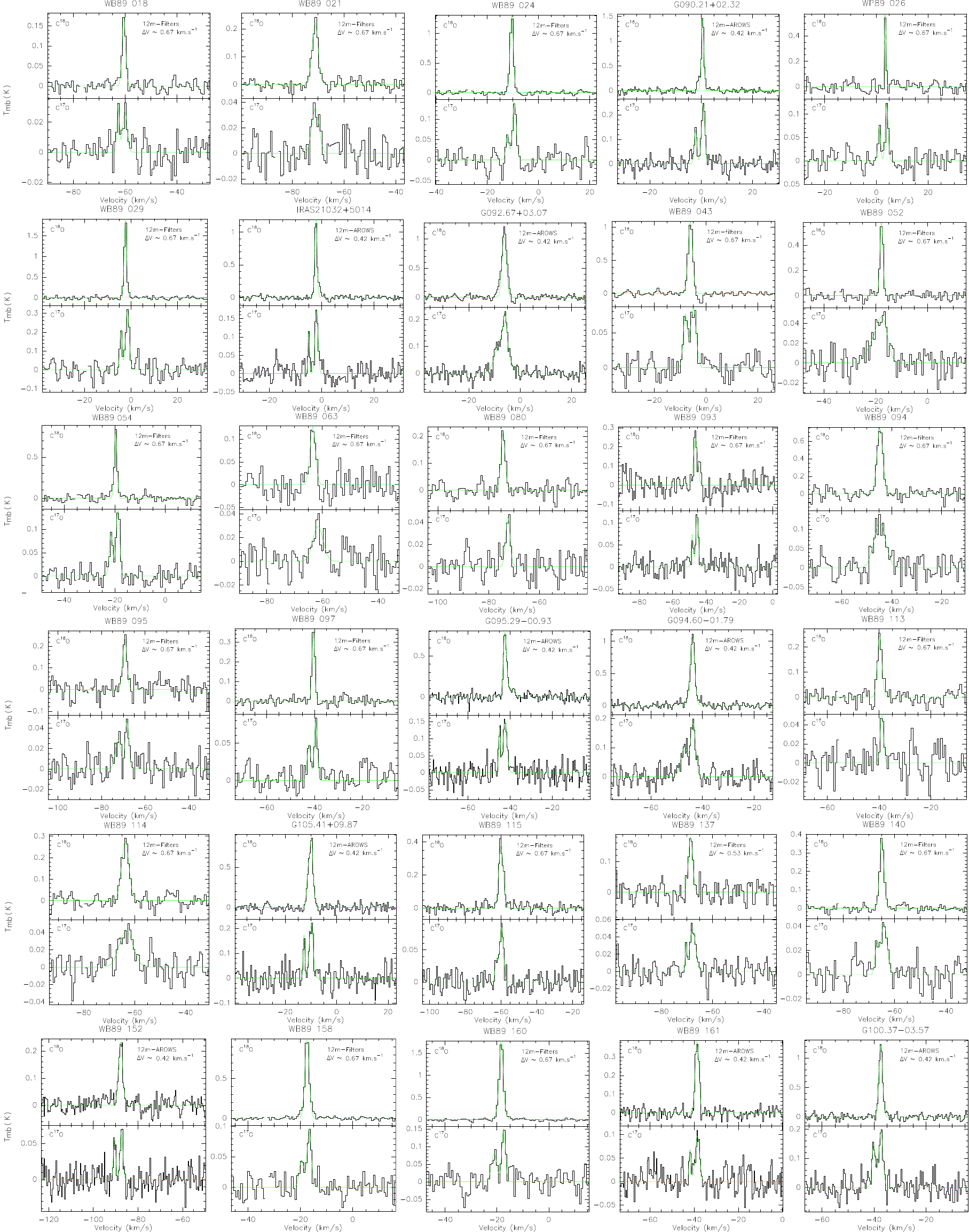}
\caption{continued}
\end{figure}

\setcounter{figure}{0}
\begin{figure}
\plotone{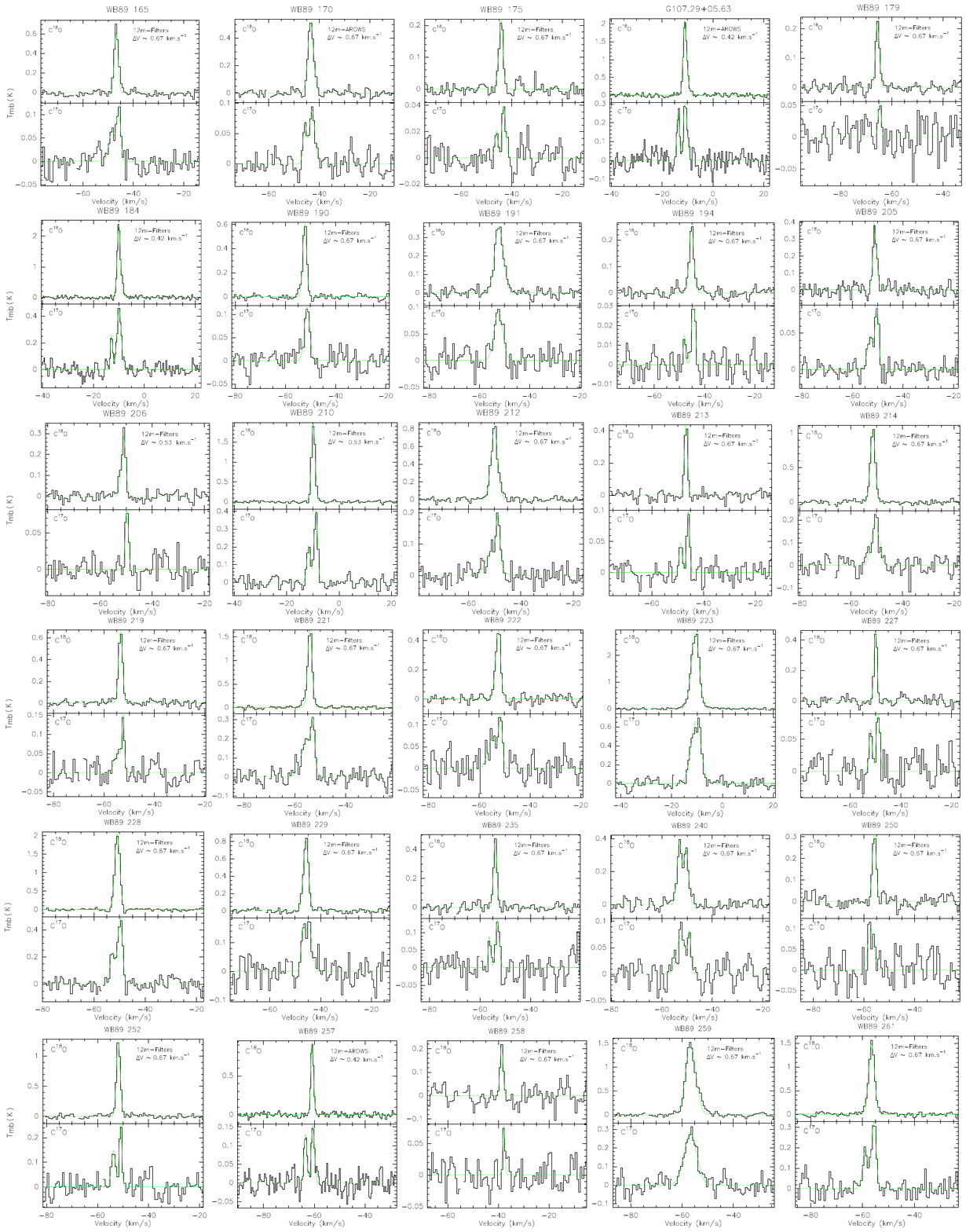}
\caption{continued}
\end{figure}

\setcounter{figure}{0}
\begin{figure}
\plotone{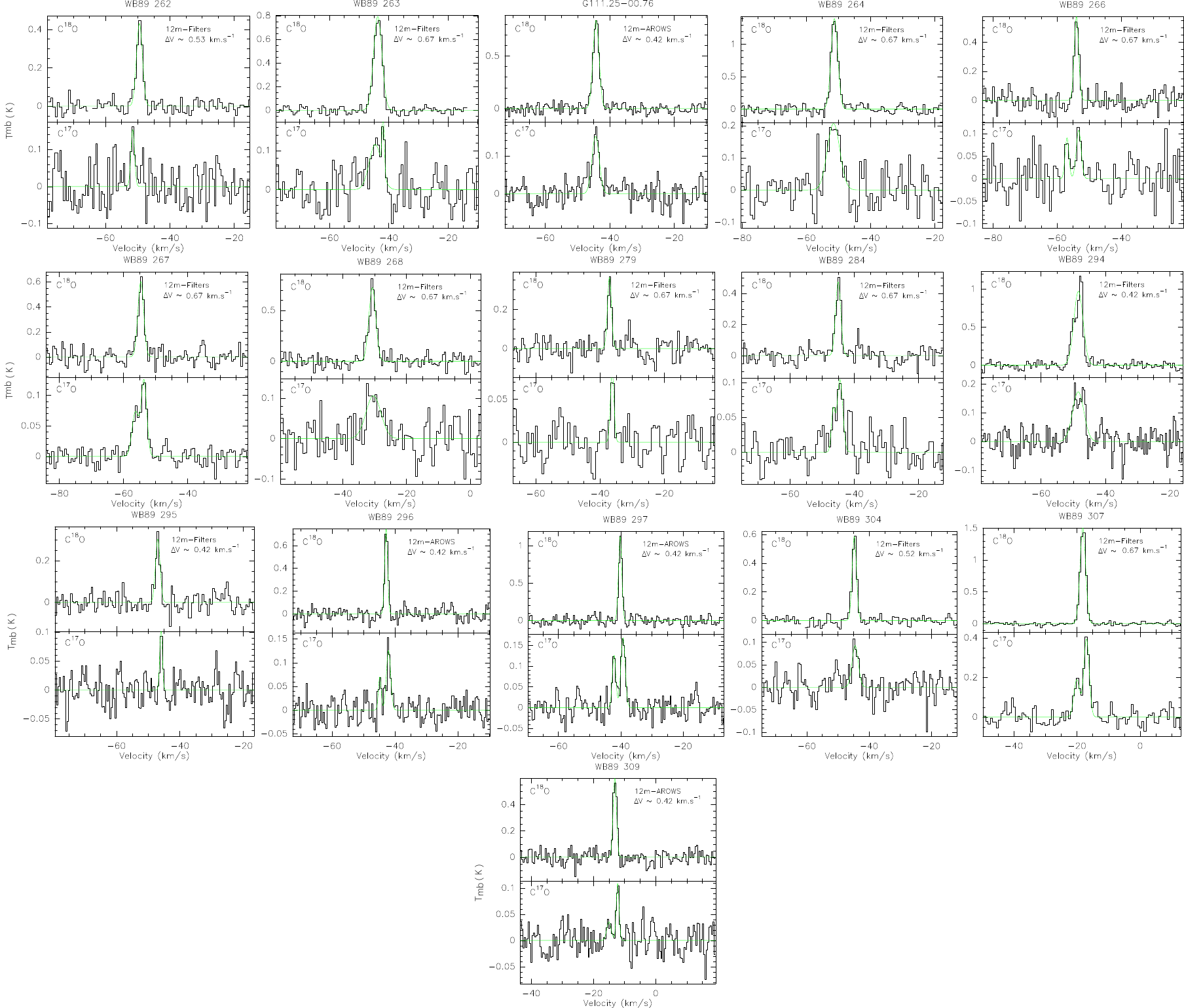}
\caption{continued}
\end{figure}

\setcounter{figure}{1}
\begin{figure}
\plotone{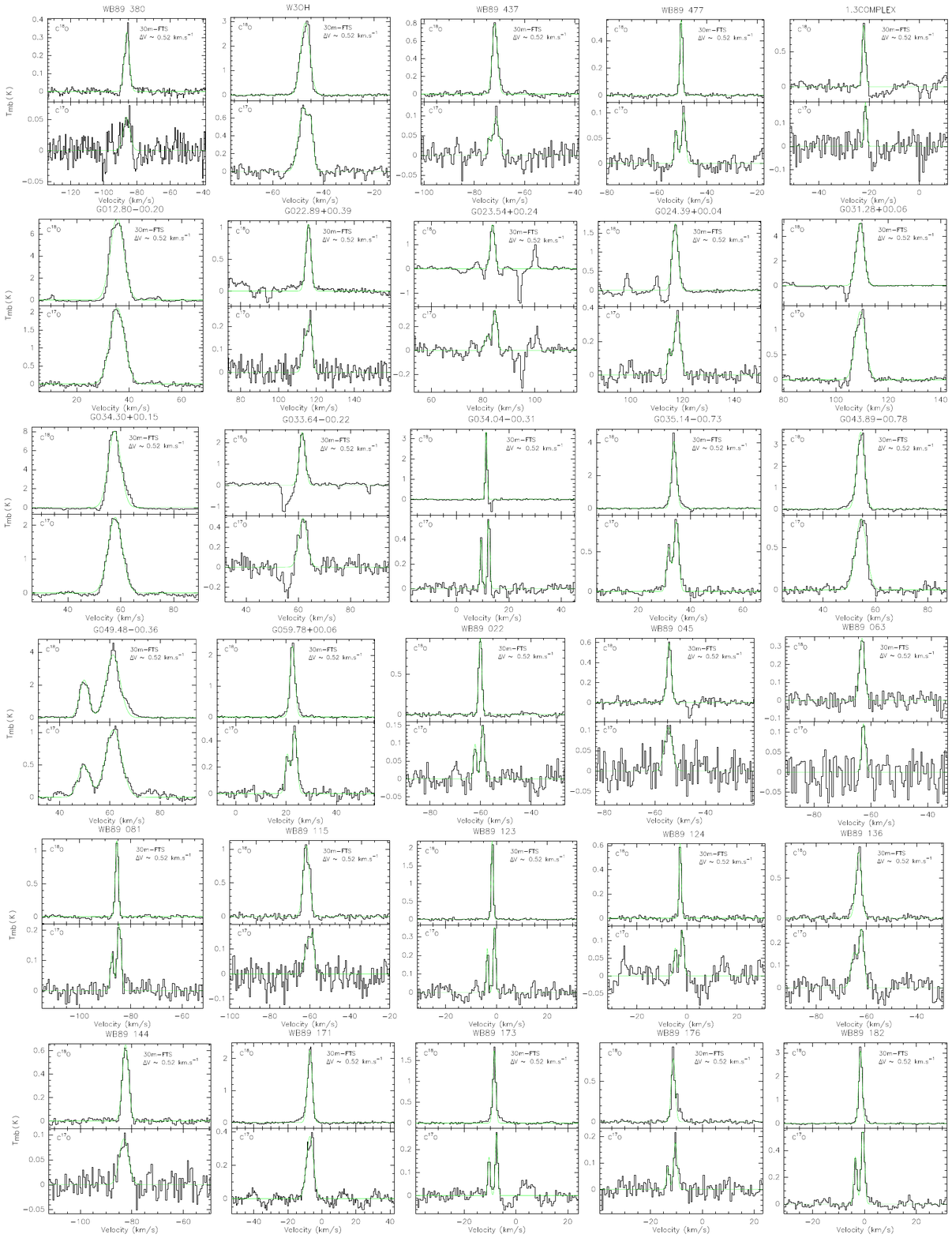}
\plotone{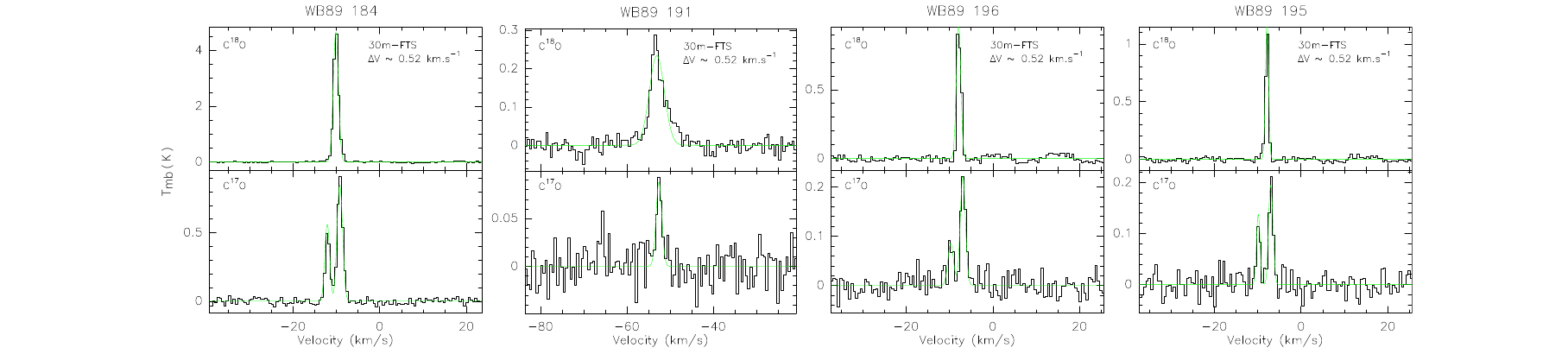}
\caption{The IRAM 30m spectra of C$^{18}$O (upper panels) and C$^{17}$O (lower panels) with green fit lines of the 34 sources in both isotopomers.}
\end{figure}

\setcounter{figure}{2}
\begin{figure}
\centering
\includegraphics[scale=1.5]{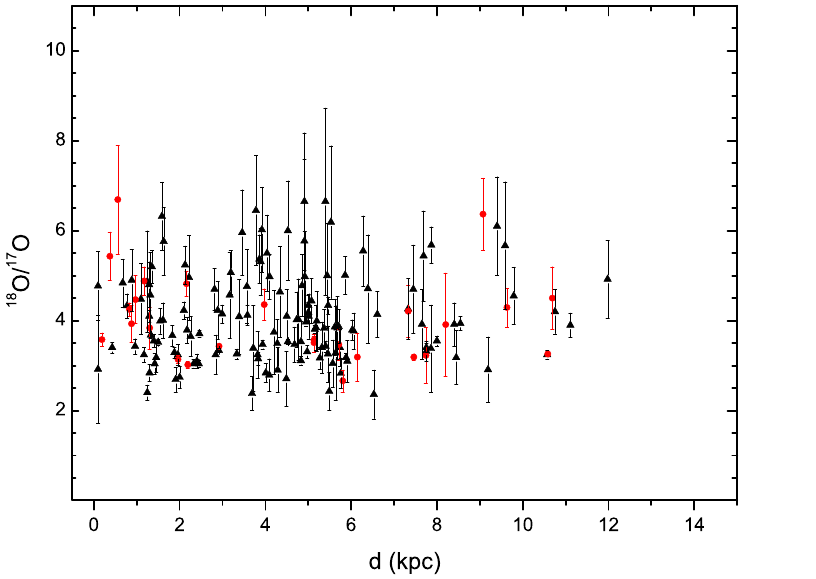}
\caption{Isotopic ratio $^{18}$O/$^{17}$O against the heliocentric distance and no significant variation can be found between them.}
\end{figure}

\setcounter{figure}{4}
\begin{figure}
\centering
\includegraphics[scale=0.8]{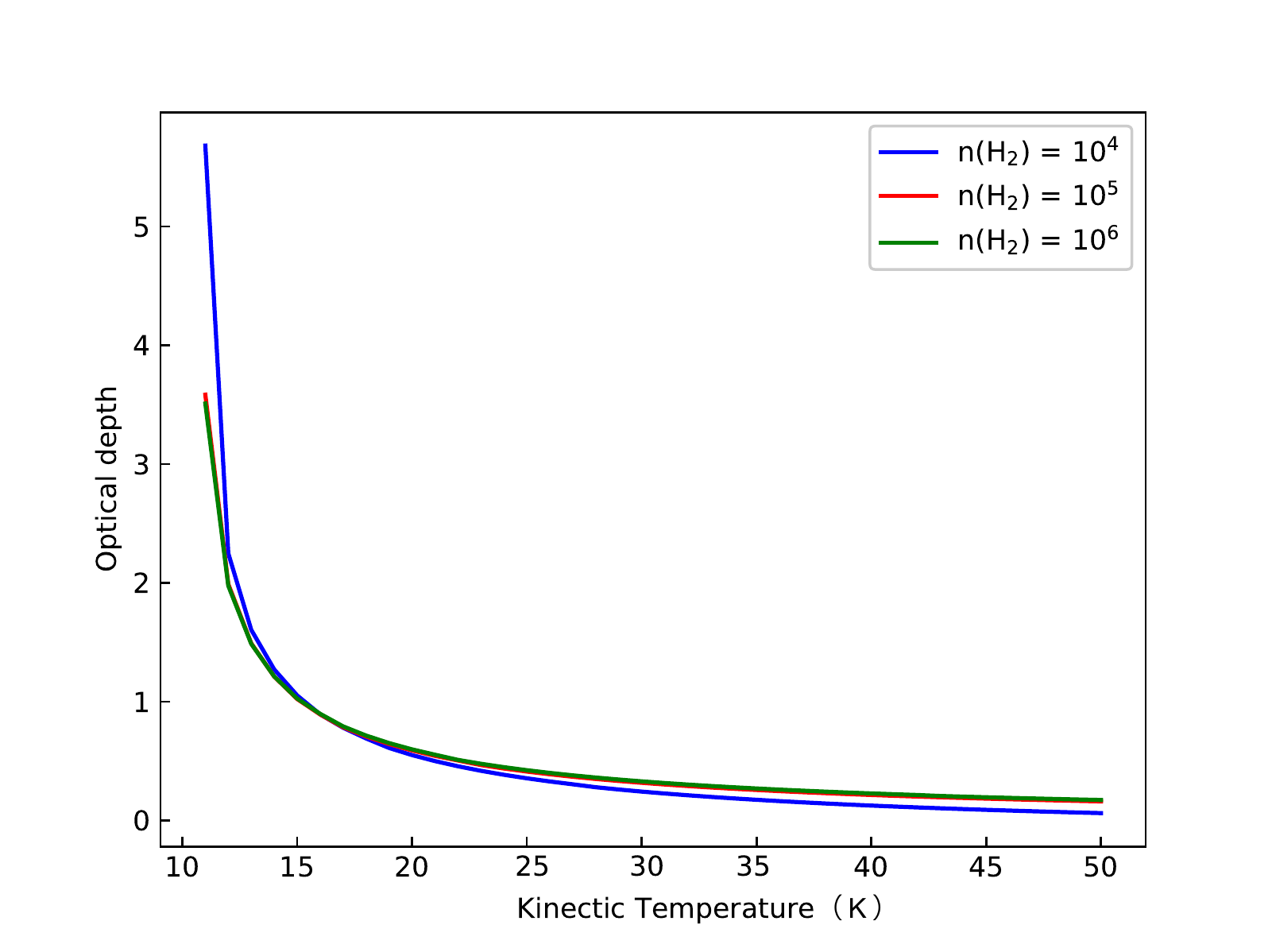}
\caption{Results of the RADEX non-LTE model calculations: the optical depth of C$^{18}$O against the kinetic temperature of the strongest source G034.30+00.15 among our sample, under different density circumstance.}
\end{figure}

\setcounter{figure}{3}
\begin{figure}
\plotone{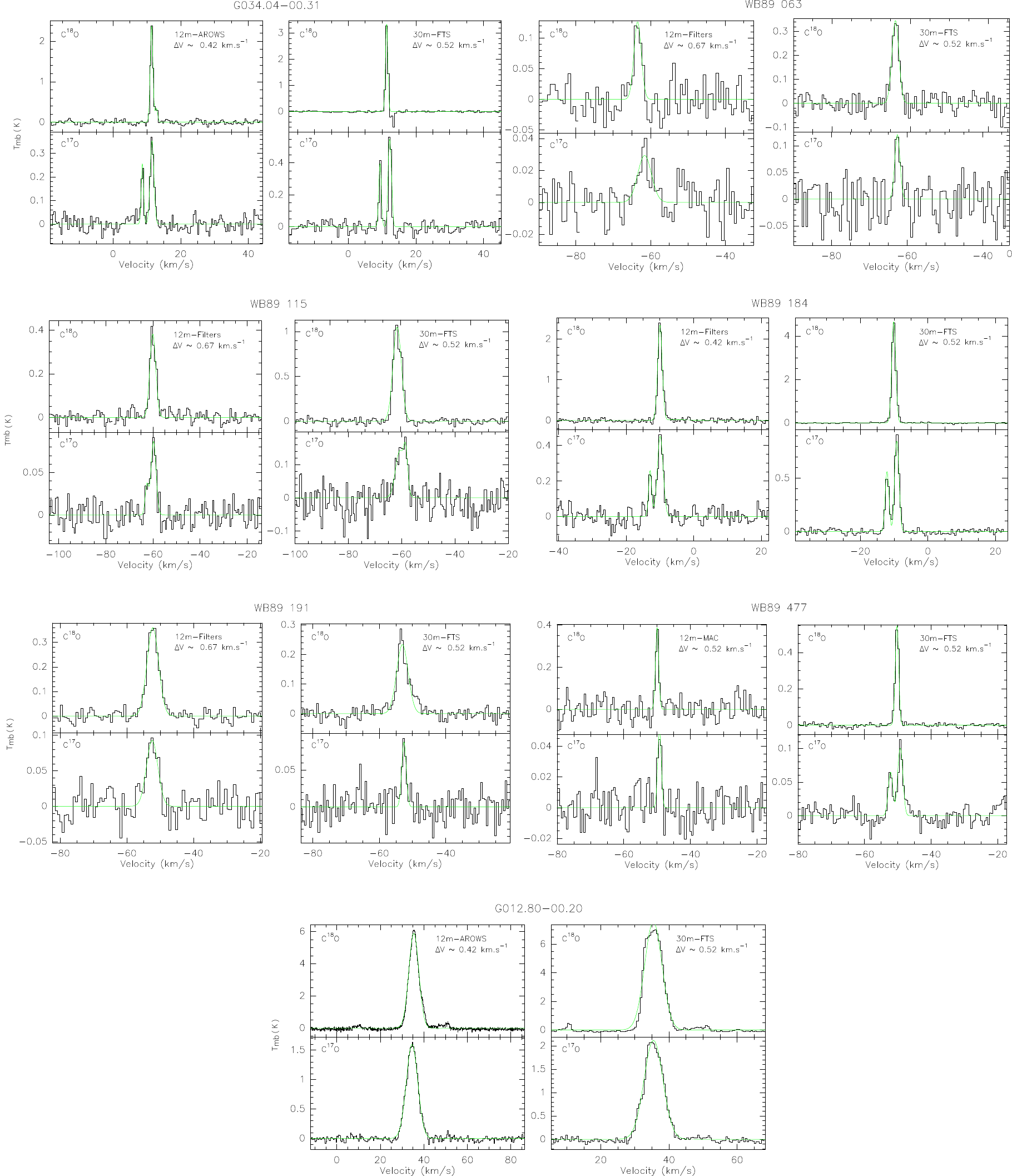}
\caption{The spectra of $ \rm{C}^{18}\rm{O}$ (upper panels) and $ \rm{C}^{17}\rm{O}$ (lower panels) with green fit lines of those 7 sources, detected by both ARO\,12M (left columns) and IRAM\,30M telescope (right columns).}
\end{figure}

\setcounter{figure}{5}
\begin{figure}
\centering
\includegraphics[scale=1.39]{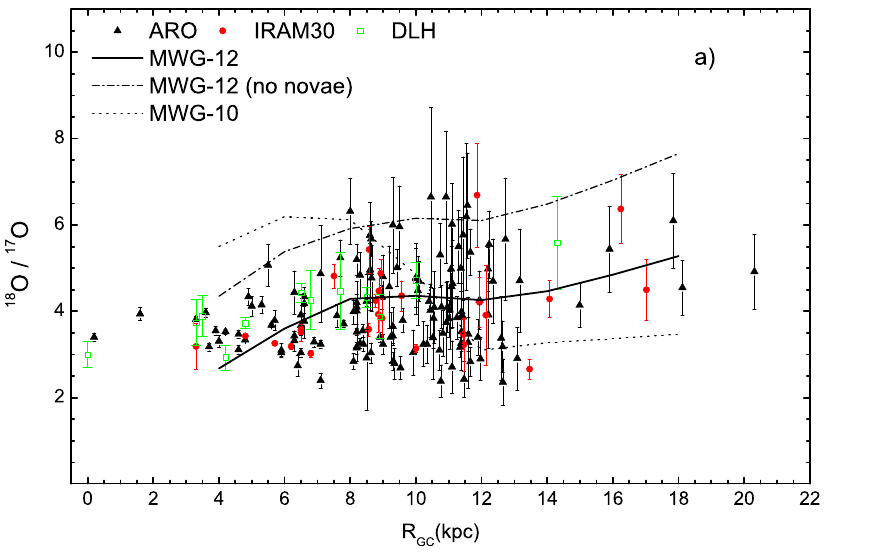}
\includegraphics[scale=1.35]{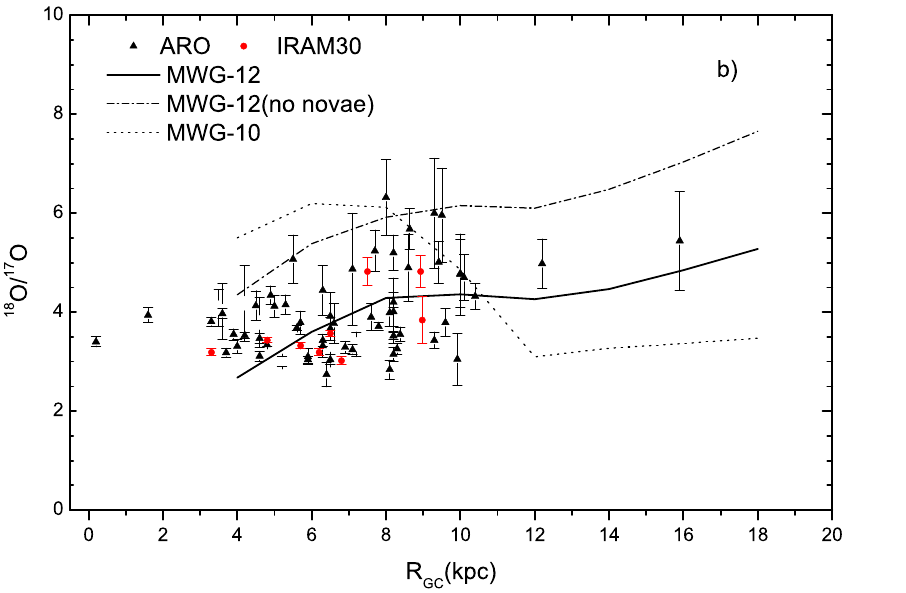}
\caption{The isotopic ratio $^{18}$O/$^{17}$O results are plotted as a function of the galactocentric distance. Upper panel for the whole sample
(Figure\,6a): the filled black triangles and red circles are the results from our ARO\,12m and IRAM\,30m measurements, respectively. Lower panel for the HMSFR sample alone (with accurate distance values by the trigonometric parallax method, Figure\,6b): same symbols as Figure\,6a. Previous Delingha\,13.7m measurements (Li et al. 2016) in green empty squares are also presented. The curves represent predictions of the newest galactic chemical evolution model: the dot curves for the model adopting the new yields by Limongi \& Chieffi (2018) for non-rotating stars (MWG-10), the dash-dot curves and the solid curves for rotating stars without or with novae, respectively (MWG-12, see details in Romano et al. 2019).}
\end{figure}

\setcounter{figure}{6}
\begin{figure}
\centering
\includegraphics[scale=1.5]{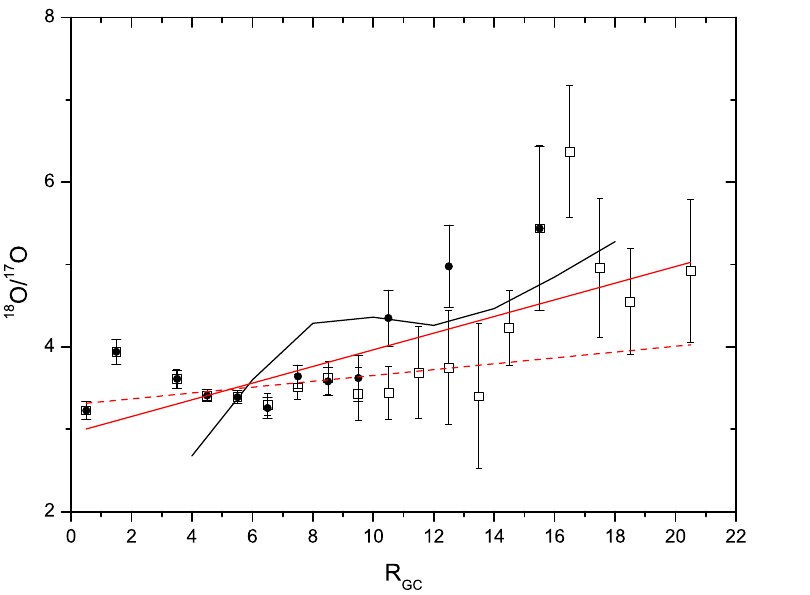}
\caption{Weighted averages of the isotopic ratio $^{18}$O/$^{17}$O in bins of 1\,kpc in R$_{GC}$ as a function of galactocentric distance. Squares and circles for the whole sample and the sub-sample of HMSFR from Reid et al. (2014), respectively. The solid and the dashed lines represent its unweighted and weighted linear fit, respectively. As in Figure 6, the curve represents predictions of the galactic chemical evolution model including both rotating stars and novae (MWG-12 in Romano et al. 2019).}
\end{figure}

\clearpage


\begin{deluxetable}{ccccccccccc}

\tablecaption{$^{18}$O/$^{17}$O isotopic ratios of large sample of molecular clouds: J=1-0 C$^{18}$O and C$^{17}$O observations}

\tablehead{
\colhead{Source Name} & \colhead{R.A.} & \colhead{DEC.} & \colhead{$R_{GC}$} & \colhead{$d$} & \colhead{Telescope} & \colhead{Line} & \colhead{rms} &\colhead{$ \int{T_{\rm mb}{\rm{d}} v}$}  & \colhead{Tpeak} & \colhead{$Ratio_{cor}$}  \\
                      &  \multicolumn{2}{c}{(J2000)}    & (kpc)              &    (kpc)  &          &         &   (${\rm K}$) &  (${\rm K} \, {\rm km} \, {\rm s}^{-1}$) & (${\rm K}$) &    }
\startdata
WB89 310	&	00:00:42.8	&	68:39:51.0	&	8.86	&	0.73	&	ARO	&	$ \rm{C}^{18}\rm{O}$	&	0.051	&		   &		&   \\
	        &		    &		    &		    &		    &		&	$ \rm{C}^{17}\rm{O}$	&		    &		   &		&	    	\\
WB89 312	&	00:02:41.3	&	64:34:04.0	&	11.39	&	4.60	    &	ARO	&	$ \rm{C}^{18}\rm{O}$	&	0.025	&		   &		&  	 \\
	        &		    &		    &		    &		    &		&	$ \rm{C}^{17}\rm{O}$	&	0.020	&		   &		&        	\\
WB89 318	&	00:07:24.6	&	64:58:02.0	&	13.66	&	7.41	&	ARO	&	$ \rm{C}^{18}\rm{O}$	&	0.045	&		   &		&  	 \\
	        & 		    &		    &		    &		    &		&	$ \rm{C}^{17}\rm{O}$	&		    &		   &		&	    	\\
WB89 322	&	00:10:39.0	&	63:46:15.0	&	10.70	&	6.16	&	ARO	&	$ \rm{C}^{18}\rm{O}$	&	0.055	&		   &		&	 \\
	        &		    &		    &		    &		    &		&	$ \rm{C}^{17}\rm{O}$	&		    &	       &		&			\\
WB89 325    & 00:14:27.0   & 64:28:30.4     & 10.56     & 3.39      & ARO   &   $ \rm{C}^{18}\rm{O}$ &0.023 &1.77(0.04)&0.85 &4.09 (0.83) \\
            &              &                &           &           &       &   $ \rm{C}^{17}\rm{O}$ & 0.028 &0.45(0.07)& 0.15 &                   \\
WB89 326    & 00:15:29.1   &  61:14:41.0    & 10.76     & 3.69      & ARO   &   $ \rm{C}^{18}\rm{O}$ &0.021 &1.08(0.04)&0.73 &2.38 (0.38) \\
            &              &                &           &           &       &   $ \rm{C}^{17}\rm{O}$ & 0.018 &0.48(0.05)& 0.14 & \\
WB89 330	&	00:20:58.0	&	62:40:18.0	&	11.60	&	4.68	&	ARO	&	$ \rm{C}^{18}\rm{O}$	&0.052	&0.73(0.02)&0.68 &3.47 (0.39) \\
	        &		    &	&		&		    &		&	$ \rm{C}^{17}\rm{O}$	&	0.027	&0.22(0.02)&	0.13 &		                 \\
\enddata
\tablecomments{Sources with blank ratio values stand for those with its
    C$^{17}$O was not detected or not observed due to its weak C$^{18}$O
    signal. Sources in italics for those with absorption features close to
    the emission in the spectra. a) For the Galactic center source 1.3
    Complex, its broad component ($\sim$50--80\,km\,s$^{-1}$) was not
    detected and the results given here are for one detected strong narrow
    component ($\sim$-22\,km\,s$^{-1}$, not used for our analysis), which
    should arise from the local gas, due to its smaller linewidth and larger
    abundance ratio than typical GC sources (Zhang et al. 2015). b) For
    those sources with two velocity components, the abundance ratio is given
    separately. c) For those 6 sources with low S/N ($\leq$3) C$^{17}$O
    spectra, their abundance ratios were not given. Column(1): source name;
    Column(2) and (3): equatorial coordinates in J2000;  Column(4): the
    galactocentric distance $R_{GC}$; Column(5): the heliocentric distance
    $d$; Column(6): used telescopes; Column(7): molecular species;
    Column(8): the Root-Mean-Square ($\rm {rms}$) value in T$_{mb}$.
    Column(9): the integrated line intensity of $ \rm{C}^{18}\rm{O}$ and $
    \rm{C}^{17}\rm{O}$, with its error in parentheses; Column(10): the line
    peak values, in T$_{mb}$; Column(11): the frequency-corrected abundance
    ratio with its error in parentheses. \\
    Table 1 is published in its entirety in the machine-readable format.
    A portion is shown here for guidance regarding its form and content.}
\end{deluxetable}

\begin{deluxetable}{ccccccccc}

\tablecaption{Comparisons of ARO\,12m and IRAM\,30m observation results for 7 sources with both $ \rm{C}^{18}\rm{O}$ and $ \rm{C}^{17}\rm{O}$ detections}

\tablehead{
\colhead{Source Name} & \colhead{Telescope} & \colhead{Line} & \colhead{rms} & \colhead{$ \int{T_{\rm mb}{\rm{d}} v}$} &\colhead{$T_{\rm peak}$}  &
\colhead{$R_{GC}$} & \colhead{$d$} & \colhead{$Ratio_{cor}$}                                                                             \\
                      &             &             & (${\rm K}$) & (${\rm K} \, {\rm km} \, {\rm s}^{-1}$)& (${\rm K}$)  &         &         &
 }
\decimalcolnumbers
\startdata
G012.80-00.20 &	ARO	&	$ \rm{C}^{18}\rm{O}$	&		0.103 	&	33.49   (	0.18 	)&	5.85    &    4.80  & 2.92 &	3.30 (	0.05 )	 \\
	          &		&	$ \rm{C}^{17}\rm{O}$	&		0.032 	&	10.62 	(	0.08 	)&	1.60 	&		   &	  & 				 \\
   	          &	IRAM&	$ \rm{C}^{18}\rm{O}$	&	 	0.105 	&	48.47 	(	0.22 	)&	7.35 	&	 	   &	  & 3.43 (	0.05 )	 \\
              &		&	$ \rm{C}^{17}\rm{O}$	&		0.052 	&	14.79 	(	0.10 	)&	2.14 	&		   &      &      	         \\
G034.04-00.31 &	ARO	&	$ \rm{C}^{18}\rm{O}$	&		0.180 	&	2.47    (	0.05 	)&	2.45    &    7.70  & 2.12 &	3.54 (	0.41 )	 \\
	          &		&	$ \rm{C}^{17}\rm{O}$	&		0.042 	&	0.73 	(	0.07 	)&	0.37 	&		   &	  &				 \\
   	          &	IRAM&	$ \rm{C}^{18}\rm{O}$	&	 	0.018 	&	2.98 	(	0.06 	)&	3.35 	&	 	   &	  & 3.33 (	0.21 )	 \\
              &		&	$ \rm{C}^{17}\rm{O}$	&		0.024 	&	1.20 	(	0.03 	)&	0.58 	&		   &      &      	     \\
WB89 063	  &	ARO	&	$ \rm{C}^{18}\rm{O}$	&		0.021 	&	0.32 	(	0.05 	)&	0.13 	&	12.13  & 8.20 &	3.42 ( 0.17 )	\\
	          &		&	$ \rm{C}^{17}\rm{O}$	&		0.008 	&	0.10 	(	0.02 	)&	0.03 	&		   &      &   			\\
	          &	IRAM&	$ \rm{C}^{18}\rm{O}$	&	 	0.027 	&	0.80 	(	0.05 	)&	0.35 	&	       &	  & 3.91 (	0.61 )	 \\
	          &		&	$ \rm{C}^{17}\rm{O}$	&		0.029 	&	0.21 	(	0.05 	)&	0.12 	&		   &	  &			    \\
WB89 115      &	ARO	&	$ \rm{C}^{18}\rm{O}$	&		0.018 	&	1.19 	(	0.05 	)&	0.40 	&	11.94  & 7.33 &	5.00 (	0.24 )	\\
	          &		&	$ \rm{C}^{17}\rm{O}$	&		0.010 	&	0.25 	(	0.03 	)&	0.08 	&		   &      &      	    \\
	          &	IRAM&	$ \rm{C}^{18}\rm{O}$	&		0.022 	&	3.29 	(	0.05 	)&	1.05 	&	       &	  & 4.37 (	0.50 )	 \\
	          &		&	$ \rm{C}^{17}\rm{O}$	&		0.033 	&	0.79 	(	0.10 	)&	0.24 	&		   &	  &	        	 \\
WB89 184  	  &	ARO	&	$ \rm{C}^{18}\rm{O}$	&		0.033 	&	3.93 	(	0.05 	)&	2.45 	&	8.93   & 1.17 &	3.34 ( 0.40	)	\\
	          &		&	$ \rm{C}^{17}\rm{O}$	&		0.032 	&	1.23 	(	0.13 	)&	0.47 	&		   &	  &				 \\
	          &	IRAM&	$ \rm{C}^{18}\rm{O}$	&		0.016 	&	6.90 	(	0.02 	)&	4.56 	&	       &	  & 3.26(	0.26 )	 \\
	          &		&	$ \rm{C}^{17}\rm{O}$	&		0.018 	&	2.22 	(	0.14 	)&	0.90 	&		   &	  &		         \\
WB89 191	  &	ARO	&	$ \rm{C}^{18}\rm{O}$	&		0.017 	&	1.58 	(	0.04 	)&	0.38 	&	11.44  & 5.71 &	3.87 (	0.38 )	 \\
	          &		&	$ \rm{C}^{17}\rm{O}$	&		0.016 	&	0.41 	(	0.04 	)&	0.12 	&		   &	  &				 \\
	          &	IRAM&	$ \rm{C}^{18}\rm{O}$	&		0.013 	&	1.00 	(	0.04 	)&	0.29 	&	       &	  & 3.61(	0.35 )	 \\
	          &		&	$ \rm{C}^{17}\rm{O}$    &       0.014   &	0.29 	(	0.02 	)&	0.09 	&		   &	  &				 \\
WB89 477	  &	ARO	&	$ \rm{C}^{18}\rm{O}$	&		0.037 	&	0.47 	(	0.05 	)&	0.38 	&	13.46  & 5.80 &	...        	 \\
	          &		&	$ \rm{C}^{17}\rm{O}$$^{a)}$	&	0.009 	&	0.07 	(	0.01 	)&	0.04 	&	       &	  &    			 \\
	          &	IRAM&	$ \rm{C}^{18}\rm{O}$	&	 	0.009 	&	0.72 	(	0.01 	)&	0.54 	&	       &	  & 2.36 (	0.31)	 \\
	          &		&	$ \rm{C}^{17}\rm{O}$	&		0.014 	&	0.32 	(	0.04 	)&	0.10 	&		   &	  &				    \\
\enddata
\tablecomments{Column(1): source name;  Column(2): used telescope;  Column(3): molecular species; Column(4): the Root-Mean-Square
($\rm {rms}$) value in T$_{mb}$ for IRAM\,30 data; Column(5): the integrated line intensity of $\rm{C}^{18}\rm{O}$ and
$\rm{C}^{17}\rm{O}$, with its error in parentheses; Column(6): the peak values of lines, as $\rm {rms}$ in T$_{mb}$;
Column(7): the galacocentric distance $R_{GC}$;
Column(8): the heliocentric distance $d$; Column (9): the frequency-corrected abundance ratio with its error in parentheses.
a): Due to its narrow ARO C$^{17}$O spectra without hfs features, the ARO ratio result of WB89\,477 is not taken.}

\end{deluxetable}

\clearpage




%
%


\end{document}